%% file: main.tex
\documentclass[runningheads]{llncs}
\usepackage[T1]{fontenc}
\usepackage{graphicx}
\usepackage{color}
\usepackage{mathpartir}
\usepackage{hyperref}
\usepackage{amsmath}

\usepackage[dvipsnames]{xcolor}
\usepackage{listings}
\usepackage{algorithm}
\usepackage[noend]{algpseudocode}

\usepackage[symbol]{footmisc}
\usepackage{marvosym} 

\definecolor{codegreen}{rgb}{0,0.6,0}  
\lstdefinelanguage{diff}{  
  morecomment=[f][\color{blue}]{@@},     
  morecomment=[f][\color{red}]-,         
  morecomment=[f][\color{codegreen}]+,       
  morecomment=[f][\color{red}]{---}, 
  morecomment=[f][\color{codegreen}]{+++},
}

\usepackage{color,appendix}
\usepackage{multirow}
\usepackage{multicol, cuted}
\usepackage[switch]{lineno}

\newcommand{\lst}{\lstinline}
\newcommand{\tf}{\texttt}
\lstdefinelanguage{Coq}%
  {morekeywords={Variable,Inductive,CoInductive,Fixpoint,CoFixpoint,%
      Definition,Lemma,Theorem,Axiom,Goal,Save,Grammar,Syntax,Intro,%
      Trivial,Qed,Intros,Symmetry,Simpl,Rewrite,Apply,Elim,Assumption,%
      Left,Cut,Case,Auto,Unfold,Exact,Right,Hypothesis,Pattern,Destruct,%
      Constructor,Defined,Fix,Record,Proof,Induction,Hints,Exists,let,in,%
      Parameter,Split,Red,Reflexivity,Transitivity,then,else,Opaque,%
      Transparent,Inversion,Absurd,Generalize,Mutual,Cases,of,end,Analyze,%
      AutoRewrite,Functional,Scheme,params,Refine,using,Discriminate,Try,%
      Require,Ensure,Load,Import,Scope,Open,Section,End,match,with,Ltac,%
      Instance,Class,With,unsigned,long,%
      Notation, fun, forall, exists, Example,%
      as,do,Inv,for,if,__u32,%
      Let,int,char,NULL,SEC,struct,void,%
	},%
   sensitive, %
   morecomment=[n]{(*}{*)},%
   morestring=[d]",%
   literate=
   {>->}{{$\rightarrowtail$}}2
   {->}{{$\rightarrow$}}1
   {<-}{{$\gets$}}1
   {==>}{==>}1
   {=>}{{$\Rightarrow$}}1
   {<->}{{$\leftrightarrow$}}2
   {|--}{{$\vdash$}}1
   {odot}{$\odot$}1 {Odot}{$\bigodot$}1
   {otimes}{$\otimes$}1 {Otimes}{$\bigotimes$}1
   {oplus}{$\oplus$}1  {Oplus}{$\bigoplus$}1
   {inx}{$\in$}1
   {star}{{$\star$}}1
   {mapsto}{{$\mapsto$}}1
   {/\\}{{$\wedge$}}1
   {union}{$\cup$}1
   {nvdash}{{$\nvdash$}}1
   {sigma}{{$\tt{\Sigma}$}}1
   {emptyset}{{$\tt{\emptyset}$}}1
   {notx}{{$\neg$}}1
   {notinx}{{$\notin$}}1
  }[keywords,comments,strings]%

\begin{document}
\title{A Formal Framework for Naturally Specifying and Verifying Sequential Algorithms}    
\titlerunning{A Formal Framework for Naturally Specifying and Verifying Algorithms}
%
\author{Chengxi Yang\inst{1}\textsuperscript{*} \and
Shushu Wu\inst{2}\textsuperscript{*} \and
Qinxiang Cao\inst{2}\textsuperscript{(\Letter)}
}

\footnotetext{\textsuperscript{*}These authors contributed equally to this work.}
\renewcommand{\thefootnote}{\arabic{footnote}}
\authorrunning{C. Yang et al.}
%
\institute{Zhiyuan College, Shanghai Jiao Tong University, Shanghai, China \and
Shanghai Jiao Tong University, Shanghai, China\\
\email{\{arcadia-y, Ciel77, caoqinxiang\}@sjtu.edu.cn}}
\maketitle              
\begin{abstract}

Current approaches for formal verification of algorithms face important limitations. For specification, they cannot express algorithms naturally and concisely, especially for algorithms with states and flexible control flow. For verification, formal proof based on Hoare logic cannot reflect the logical structure of natural proof.
To address these challenges, we introduce a formal framework for naturally specifying and verifying sequential algorithms in Coq. We use the state relation monad to integrate Coq's expressive type system with the flexible control flow of imperative languages. It supports nondeterministic operations and customizable program states, enabling specifying algorithms at an appropriate level of abstraction.
For verification, we build a Hoare logic for the monad and propose a novel two-stage proof approach that separates natural logical reasoning from mechanical composition. It reflects the logical structure of natural proof, enhancing modularity and readability.
We evaluate the framework by formalizing the Depth-First Search (DFS) algorithm and verifying the Knuth-Morris-Pratt (KMP) algorithm.

\keywords{Formal Verification \and Monad \and Hoare Logic \and Coq.}
\end{abstract}

\input{introduction}

\input{monad}

\input{logic}

\input{related}

\input{conclucion}

\begin{credits}
\subsubsection{\ackname} This material is based upon work supported by NSF China 62472274  and 92370201.

\subsubsection{\discintname}
The authors have no competing interests to declare that are
relevant to the content of this article.
\end{credits}

\newpage
\bibliographystyle{splncs04}
\bibliography{bibliography}

\newpage
\appendix
\input{appendix}

\end{document}

%% file: introduction.tex
\section{Introduction}\label{sec::intro}

The formal verification of algorithms aims to formally specify algorithms and mathematically state and prove their functional correctness. 
It is canonical to formalize sequential algorithms in proof assistants like Coq
\footnote{Currently Coq was recently renamed to ``Rocq'', more details are available at \url{https://rocq-prover.org/about\#Name}. However, we still use the name ``Coq'' in this paper since we adopt an older version (8.15.2) for formalization.}
\cite{Coq} and Isabelle/HOL \cite{Isabelle}, and there have been various approaches.  

A common approach is to define algorithms as pure functions and prove their correctness using proof assistants' built-in logic, such as Verified Functional Algorithms \cite{Appel:SF3}.
Still, many algorithms are naturally expressed in an imperative sequential form rather than pure functional form. Moreover, the restriction to structural recursion in Coq further complicates the task of specifying algorithms. For instance, even simple graph algorithms like Depth-First Search (DFS) become cumbersome to define under these constraints.

Another approach involves defining a simple imperative language, such as Imp \cite{Pierce:SF2}, and formalizing algorithms within it. While this method allows for imperative constructs, it fails to leverage Coq's powerful type system, limiting its expressiveness for abstract operations like selecting an arbitrary element from a set.

In addition, the Isabelle Refinement Framework \cite{Refine_Monadic-AFP} enables users to formulate nondeterministic algorithms in a monadic
style. Nevertheless, since it only supports functional and stateless programs, it is not convenient when formulating stateful algorithms (e.g. algorithms with a working set or state machine).

Beyond formalization, current verification approaches face additional limitations. Most frameworks for imperative algorithm verification rely on Hoare logic \cite{hoare}, which structures proofs according to the program's syntactic structure. This often requires grouping all propositions about the current program state into a conjunction and applying Hoare rules corresponding to the current statement. In contrast, natural proofs tend to organize propositions based on their logical relationships, following the proof's natural structure rather than the program's. This discrepancy becomes particularly evident when propositions span different program segments or loops, which Hoare logic struggles to handle elegantly.

Consider, for example, the \textit{match} procedure in the Knuth–Morris–Pratt (KMP) algorithm \cite{KMP}, which finds the first occurrence of a pattern string in a text string. The procedure is shown in Algorithm~\ref{match}.

\begin{algorithm}
\caption{Match procedure in the KMP algorithm}\label{match}
\begin{algorithmic}[1]
\Procedure{Match}{$patn$, $text$, $next$}
\State {$j = 0$}
\For{$i$ \textbf{from} $0$ \textbf{to} $text.len$}
    \State {$ch = text[i]$}
    \Loop
        \If {$patn[j] = ch$}
            \State $j \gets j + 1$
            \State \textbf{break}
        \EndIf
        \If {$j = 0$}
            \State \textbf{break}
        \EndIf
        \State $j \gets next[j-1]$
    \EndLoop
    \If {$j = patn.len$}
        \State \Return $i - patn.len + 1$
    \EndIf
\EndFor
\State \Return $-1$
\EndProcedure
\end{algorithmic}
\end{algorithm}

In Algorithm~\ref{match}, $patn$ represents the pattern string to be located within the text string $text$, while $next$ is an array containing shift information critical to the algorithm's efficiency. It is precomputed by the table-building procedure in the KMP algorithm, and it represents the prefix function\footnote{More details of the prefix function can be found in classic algorithm textbooks \cite{introalgo}.} of $patn$, which stores for each position $j$ in $patn$ the length of the longest proper prefix of $patn[0..j]$ that is also a suffix. $a.len$ denotes the length of an array $a$.

To verify the procedure using the Hoare logic, one usually first provides a loop invariant $I$ for the for-loop such as $\textit{jrange} \land \textit{partial\_match} \land \textit{partial\_bound} \land \textit{no\_occur}$, where
\begin{itemize}
    \item \textit{jrange} means that $0 \leq j < next.len$ so $j$ is a valid index for $next$ and $patn$.
    \item \textit{partial\_match} means that $j$ is a partial match result for $text[0..i]$\footnote{$\textit{array}[s..t]$ refers to the 0-indexed segment of $\textit{array}$ ranging from $s$ (included) to $t$ (excluded).}, i.e. $patn[0..j] = text[i-j..i]$.
    \item \textit{partial\_bound} asserts that $j$ is an upperbound for the partial match result for $text[0..i]$. This with \textit{partial\_match} ensures $j$ is the best result.
    \item \textit{no\_occur} states that there's no occurrence of $patn$ in $text[0..i]$.
\end{itemize}
Then she aims to prove the for-loop body preserves $I$ if the loop continues.
To do so, she may assert another invariant $I'$ for the inner loop such as $\textit{jrange} \land \textit{partial\_match} \land \textit{presuffix\_inv}$, where
\textit{presuffix\_inv} states some proposition that any partial match result $k$ for $text[0..i+1]$ must obey.
Next she proves that the inner loop body preserves $I'$, $I'$ can derive some propositions $P$ when the inner loop breaks, and with $P$ the for-loop body preserves $I$.
Finally she can prove some postconditions with $I$ when the procedure returns.

In contrast, a natural proof might proceed as follows:
\textit{jrange} trivially holds throughout the for-loop due to preconditions.
Based on \textit{jrange}, the inner loop preserves \textit{partial\_match} and hence outer loop preserves \textit{partial\_match}.
Furthermore, we can prove the inner loop also preserves \textit{presuffix\_inv}. As a consequence, \textit{partial\_bound} and \textit{no\_occur} are invariants of the for-loop.
These invariants collectively lead to the desired postconditions.

As shown in the example, a natural proof tends to incrementally assert and prove properties of the program according to their logical relevance and relations. In the process, the logical dependency between propositions (e.g. \textit{no\_occur} depends on \textit{jrange}) is also naturally presented. This reveals the gap between the natural proof and the Hoare logic-based formal proof. 

To address these challenges, we present a formal framework for naturally specifying and verifying sequential algorithms in Coq. Our approach introduces a state relation monad for algorithm specification based on the denotational semantics. The monad, defined over the ternary relation of initial state, return value, and resulting state, supports imperative constructs such as general recursion and loops break. It also integrates Coq's powerful type system with nondeterministic operations, enabling users to specify algorithms at an appropriate level of abstraction and customize program states as needed.
We also provide stateless and errorful variants of the monad for different needs.

For verification, we develop a proof framework for partial correctness tailored to our monad. It includes Hoare rules for various statements and introduces a novel approach to organizing proofs. This approach divides proofs into two parts: \textit{essential proof} that captures the key logical implications in each basic block, and \textit{mechanized proof} that combines propositions to establish end-to-end correctness. The latter can be highly automated. This results in proof that is more natural, modular, and readable.

We evaluate our framework by formalizing the DFS algorithm, and specifying and verifying the KMP algorithm. In another work \cite{wu2025}, we further prove the correctness of a C program by proving it refines the KMP algorithm we specify in our framework. This highlights the real-world applicability and versatility of our framework.
The source code of our framework is available through this link: \url{https://github.com/Arcadia-Y/TASE25-Artifact}.

\paragraph{Outline.} The rest of the paper is organized as follows: Section~\ref{sec::monad} introduces the state relation monad used for defining algorithms and uses it to formalize the DFS algorithm. Section~\ref{sec::logic} presents the Hoare logic and our proof approach, with the KMP algorithm as a case study. In Section~\ref{sec::related}, we discuss related work, and in Section~\ref{sec::conclu}, we conclude with a summary and directions for future work.

%% file: monad.tex
\section{State Relation Monad}\label{sec::monad}
\subsection{Monad Design}
Our framework is based on the monad, a well-established abstraction in functional programming for structuring computations with effects \cite{Moggi91}. Its definition contains a type constructor \lstinline{M: Type -> Type} and two operators:
\begin{itemize}
    \item \lstinline{return: A -> M A} (abbreviated as \lstinline{ret}), which takes a value of type \lstinline{A} and wraps it as a value of type \lstinline{M A}.
    \item \lstinline{bind: (M A) -> (A -> M B) -> (M B)}, which takes a monadic value \lstinline{m} of type \lstinline{M A} and a function \lstinline{f} of type \lstinline{A -> M B}. It tries to somehow unwrap \lstinline{m}, applies \lstinline{f} to it and returns the result as another monadic value. 
\end{itemize}

In our framework, we model a program as a \textit{state relation monad}, defined as a ternary relation over $\tt{\Sigma \times A \times \Sigma}$, where $\tt{\Sigma}$ is the type of the program state and $\tt{A}$ is the type of the return value. This relation encodes the denotational semantics of a nondeterministic program $c$ : \ $(s_1, r, s_2) \in c$ means that, starting from the state $s_1$, program $c$ may terminate at $s_2$ and return $r$. We use the \lst{unit} type with only one value \lst{tt} to represent cases of no return value.

Our framework is based on a Coq library of sets \cite{setlib}, where sets and relations are represented as curried functions, such as \lstinline{A -> Prop} and \lstinline{A -> B -> Prop}. The definition of a monadic program is as follows\footnote{For simplicity and readability, some tedious portions of Coq code (e.g., implicit type parameters) are omitted in this paper.}:
\begin{lstlisting}
Definition program (sigma A: Type): Type := sigma -> A -> sigma -> Prop.
\end{lstlisting}

\newcommand{\bind}{\tf{bind}}
\newcommand{\ret}{\tf{ret}}

For any type \lstinline{sigma}, \lstinline{program sigma} is the type constructor for the state relation monad.
The definitions of the two basic monad operators follow directly from the denotational semantics:
\begin{itemize}
    \item For any $a$ of type \lstinline{A}, \lstinline{ret}$(a)$is value of type \lstinline{program sigma A} defined as
    \[ (s_1, r, s_2) \in \texttt{ret}(a)  \iff r = a \ \land \ s_1 = s_2 \]
    \item For any $c$ of type \lstinline{program sigma A} and $f$ of type \lstinline{A -> program sigma B},   
    \lstinline{bind}$(c,f)$ is a value of type \lstinline{program sigma B} defined as
    \[ (s_1, b, s_3) \in \texttt{bind}(c, f) \iff 
       \exists\  a \ s_2, \  (s_1, a, s_2) \in c \ \land \ (s_2, b, s_3) \in f(a) \]
\end{itemize}

Intuitively, \lstinline{ret} returns a value without changing the state, and \lstinline{bind} composes two programs by passing the return value and terminal state of the first program to the second program.
They satisfy the standard monad laws\footnote{Here, equality ($=$) denotes program equivalence, which corresponds to the equality of the underlying ternary relations, i.e. double inclusion.}:
\begin{enumerate}
    \item \lst{ret} is the left identity for \lst{bind}: $\bind (\ret(x), f) = f(x)$.
    \item \lst{ret} is the right identity for \lst{bind}: $\bind (c, \ret) = c$.
    \item \lst{bind} is associative: $\bind(\bind(c, f), g) = \bind(c, \lambda x. \bind(f(x), g)) $. 
\end{enumerate}

To achieve an imperative-style syntax, we adopt a notation
similar to Haskell's do-notation:
\begin{lstlisting}
Notation "x <- c1 ;; c2" := (bind c1 (fun x => c2)) ...
Notation "e1 ;; e2" := (bind e1 (fun _: unit => e2) ...
\end{lstlisting}

The expressiveness of the state relation monad stems from the fact that defining a program is equivalent to defining a ternary relation in Coq's logical system. Combined with customizable states, this allows users to define program statements at an appropriate level of abstraction, providing high extensibility and flexibility.

In addition to user-defined statements, we provide several operators for convenient program construction:
\begin{itemize}
    \item \lstinline{choice} stands for nondeterministic choice between two programs.
    \item \lstinline{assume} adds a logical proposition regarding the program state as an assumption. \lstinline{assume'} is the notation for assumptions independent of the state.
    \item \lstinline{any} returns an arbitrary value of a given type without changing the state.
    \item \lstinline{update} modifies the state according to a binary relation over program states.
\end{itemize}
\[\begin{aligned} 
    &\tf{ choice}(f, g) := f \cup g \\
    \forall P:\mathtt{\Sigma \to} \tf{ Prop},\ (s_1, \tf{tt}, s_2) \in &\tf{ assume}(P) \iff P(s_1) \land s_1 = s_2 \\
    \forall A:\tf{Type},\ (s_1, a, s_2) \in &\tf{ any}(A) \iff s_1 = s_2 \\
    \forall R:\mathtt{\Sigma \to \Sigma \to} \tf{ Prop},\ (s_1, \tf{tt}, s_2) \in &\tf{ update}(R) \iff R(s_1, s_2)
\end{aligned}\]

By combining \lstinline{choice} and \lstinline{assume}, we can easily express common branching statements in imperative programs. For example, the following program computes the absolute value of an integer:
\begin{lstlisting}
Example compute_abs: program unit Z :=
  choice (assume' (z >= 0);; ret z) 
         (assume' (z < 0);; ret (-z)).
\end{lstlisting}

We can also use \lstinline{any} and \lstinline{assume} to define nondeterministic abstract operations. For instance, the following program returns an arbitrary prime number:
\begin{lstlisting}
Example any_prime: program unit nat :=
  x <- any nat;;
  assume' (notx exists (m n: nat), m > 1 /\ n > 1 /\ x = m * n);;
  ret x.
\end{lstlisting}

To express recursions and loops, we follow the standard approach in the denotational semantics by using the least fixed point in the Kleene fixed-point theorem \cite{Winskel:1993}. For any directed-complete partial order $\tt A$ with a least element $\bot$ and any function $f: \texttt{A} \to \texttt{A}$, we define \lstinline{Lfix}\footnote{It is defined for any $f$, although to apply the Kleene fixed-point theorem, $f$ should be monotone and continuous.} as the supremum of the set produced by iterating $f$ on $\bot$. 
\[
\texttt{Lfix}(f) := \sup(\{f^n (\bot) \mid n \in \mathbf{N}\})
\]

Recursion can then be expressed directly using \lstinline{Lfix}. For example, the following program computes the Fibonacci number:
\begin{lstlisting}
Example Fibonacci: nat -> program unit nat :=
  Lfix
  (fun (W: nat -> program unit nat) (n: nat) => 
    choice
      (assume' (n <= 1);; ret n)
      (assume' (n > 1);;
        x <- W (n - 1);;
        y <- W (n - 2);;
        ret (x + y))).
\end{lstlisting}

We also define various loops using the least fixed point. For instance, we define loops with break to express flexible control flows. We first define an inductive type \lstinline{ContinueOrBreak} similar to a sum type to represent results with control flow annotation, and then formalize loops using \lstinline{Lfix}.
\begin{lstlisting}
Inductive ContinueOrBreak (A B: Type): Type :=
| by_continue (a: A)
| by_break (b: B).
Definition repeat_break_f
  (body: A -> program sigma (ContinueOrBreak A B)) :=
  fun (W: A -> program sigma B) (a: A) =>
      x <- body a;;
      match x with
      | by_continue a' => W a'
      | by_break b => ret b
      end.
Definition repeat_break
  (body: A -> program sigma (ContinueOrBreak A B)): A -> program sigma B :=
  Lfix (repeat_break_f body).
Definition continue (a: A): program sigma (ContinueOrBreak A B) := 
  ret (by_continue a).
Definition break (b: B): program sigma (ContinueOrBreak A B) :=
  ret (by_break b).
\end{lstlisting}

Using \lstinline{continue} and \lstinline{break}, we can construct loops with break. Below is an example of a loop that computes hailstone numbers.
\begin{lstlisting}
Example hailstone: Z -> program unit Z :=
  repeat_break
  (fun (x: Z) =>
    choice
      (assume' (x <= 1);; break x)
      (assume' (x > 1);;
       choice
        (assume' (exists k, x = 2 * k);;
        continue (x / 2))
        (assume' (exists k, x = 2 * k + 1);;
        continue (3 * x + 1)))).
\end{lstlisting}

For some algorithms, it is unnecessary to involve states; for some other algorithms, we need to model errorful computations. To address these cases, we provide stateless and errorful variants of our state relation monad: the set monad and state relation monad with error. Their syntax is very similar to the original monad. See appendix \ref{appendixA} for more details.

\subsection{Case Study: Formulating the DFS Algorithm}

We formalize the Depth-First Search (DFS) algorithm in its imperative form using the state relation monad. Our definition of directed graphs and the step relation is based on a library of formalized graph theory \cite{graphlib}.
\begin{lstlisting}
Record PreGraph (Vertex Edge: Type) := {
  vvalid : Vertex -> Prop; (* vertex set *)
  evalid : Edge -> Prop; (* edge set*)
  src : Edge -> Vertex; (* source of an edge *)
  dst : Edge -> Vertex  (* destination of an edge *)
}.
Record step_aux (pg: PreGraph V E) (e: E) (x y: V): Prop := {
  step_evalid: pg.(evalid) e;
  step_src_valid: pg.(vvalid) x;
  step_dst_valid: pg.(vvalid) y;
  step_src: pg.(src) e = x;
  step_dst: pg.(dst) e = y;
}.
Definition step (pg: PreGraph V E) (x y: V): Prop :=
  exists e, step_aux pg e x y.
\end{lstlisting}

The program state for the DFS algorithm consists of a \lstinline{visited} set that stores visited vertices and a \lstinline{stack} that maintains the search path.
\begin{lstlisting}
Record state (V: Type): Type := {
  stack: list V;
  visited: V -> Prop;
}.
\end{lstlisting}

We also define several basic operations required for DFS:
\[\begin{aligned}
    &\tf{visit}(v) :=  \\
    & \quad \tf{update}(\lambda s_1 s_2. \ s_2.\tf{visited} = s_1.\tf{visited} \cup \{v\} \ \land \ s_2.\tf{stack} = s_1.\tf{stack}) \\
    & \tf{push\_stack}(v) :=  \\
    & \quad \tf{update}(\lambda s_1 s_2. \ s_2.\tf{stack} = v::s_1.\tf{stack} \ \land \ s_2.\tf{visited} = s_1.\tf{visited}) \\
    & \tf{pop\_stack} :=  \\
    & \quad  \lambda s_1 v s_2. \ s_1.\tf{stack} = v::s_2.\tf{stack} \ \land \ s_2.\tf{visited} = s_1.\tf{visited} \\
    & \tf{if\_all\_neighbor\_visited}(pg, u) := \\
    & \quad \tf{assume}(\lambda s. \ 
     \forall v,\ \tf{step}(u, v) \to v \in s.\tf{visited})
\end{aligned}\]
\begin{itemize}
    \item $\tf{visit}(v)$ marks vertex $v$ as visited and leaves the stack unchanged.
    \item $\tf{push\_stack}(v)$ pushes vertex $v$ onto the stack without modifying the visited set.
    \item $\tf{pop\_stack}$ pops the top vertex from the stack and returns it. Note how relation enables us to define the action of modifying the stack and returning the popped value concisely.
    \item $\tf{if\_all\_neighbor\_visited}(pg, u)$ assumes that all neighbors of vertex $u$ have been visited.
\end{itemize}

The DFS algorithm is then defined as follows:
\begin{lstlisting}
Definition DFS_body (pg: PreGraph V E): V -> program (state V) unit :=
  fun u =>
    choice
     (if_all_neighbor_visited pg u;;
      choice
       (assume (fun s => s.(stack) = nil);; break tt)
       (v <- pop_stack;; continue v))
     (v <- any V;;
      assume (fun s => notx v inx s.(visited));;
      assume' (step pg u v);;
      push_stack u;;
      visit v;;
      continue v).
Definition DFS (pg: PreGraph V E): V -> program (state V) unit :=
  fun u =>
    visit u;; repeat_break (DFS_body pg) u.
\end{lstlisting}

The algorithm works as follows:
\begin{enumerate}
    \item It visits the first vertex and begins to search from it. 
    \item If all neighbors of the current vertex \lstinline{u} have been visited, it either terminates (if the stack is empty) or backtracks by popping the stack.
    \item Otherwise, it nondeterministically selects an unvisited neighbor \lstinline{v}, pushes \lstinline{u} onto the stack, marks \lstinline{v} as visited, and continues the search from \lstinline{v}.
\end{enumerate}

This formulation is concise and appropriately abstract, as it specifies neither the order in which vertices are visited, nor the concrete data structures used, aligning with the nondeterministic nature of DFS.
We proved that a vertex is visited after the DFS if and only if it is reachable from the starting vertex based on the formulation.

%% file: logic.tex
\section{Proof Framework}\label{sec::logic}
\subsection{Hoare Logic}
We develop a Hoare logic for our state relation monad to prove the partial correctness of algorithms. Drawing inspiration from Hoare Type Theory (HTT) \cite{HTT}, which integrates dependent types with Hoare triples, our logic adapts HTT’s principles to a relational semantics while addressing imperative and nondeterministic constructs. 

In our framework, a Hoare triple \lstinline{Hoare P c Q} asserts that, for any initial state satisfying \lstinline{P}, if the program \lstinline{c} terminates, then resulting state and return value satisfy \lstinline{Q}.
\begin{lstlisting}
Definition Hoare (P: sigma -> Prop) (c: program sigma A) (Q: A -> sigma -> Prop) :=
  forall s1 a s2, P s1 -> (s1, a, s2) inx c -> Q a s2.
\end{lstlisting}

We prove the following core Hoare rules\footnote{We use the canonical notation $\{P\}c\{Q\}$ to denote a Hoare triple. Besides, for simplicity, some propositions may require lifting (e.g. $P \land Q$ may mean $\lambda s. P(s) \land Q(s)$).}.
Basic rules including bind, return and consequence are adapted from HTT's typing judgements.
Choice rule enables compositional proof for branching programs.
Assume, any and update rules provide strongest postcondition for our program constructs.
For user-defined statements, step rule offers a general strongest postcondition.
Pre-exist rule is useful for extracting existential variables in the precondition, which are often introduced by previous rules.
We also adapt the conjunction rule into our logic to modularize proof for complicated postconditions, which is fundamental to our two-stage proof approach.

\begin{mathpar}
\inferrule*[left=Bind]
  {\{P\} f \{Q\} \\ \forall a, \ \{Q(a)\} g(a) \{R\}}
  {\{P\} \tf{bind}(f, g) \{R\}}

\inferrule*[left=Ret]
  { P: \tt A \to \Sigma \to \tf{Prop} }
  {\{P(a)\} \tf{ret}(a) \{P\}}

\inferrule*[left=Choice]
  {\{P\} f \{Q\} \\ \{P\} g \{Q\}}
  {\{P\} \tf{choice}(f, g) \{Q\}}

\inferrule*[left=Assume]
  { }
  {\{P\} \tf{assume}(Q) \{P \land Q\}}

\inferrule*[left=Any]
  { }
  {\{P\} \tf{any}(A) \{P\}}

\inferrule*[left=Update]
  { f: \tt \Sigma \to \Sigma \to \tf{Prop} }
  {\{P\} \tf{update}(f) \{\lambda a s_2 . \ \exists s_1, f(s_1, s_2) \land P(s_1)\}}

\inferrule*[left=Step]
  {f: \tt \Sigma \to A \to \Sigma \to \tf{Prop}}
  {\{P\} f \{\lambda a s_2 . \ \exists s_1, f(s_1, a, s_2) \land P(s_1)\}}

\inferrule*[left=Conseq]
  {P_1 \to P_2 \\ \{P_2\} f \{Q_2\} \\ Q_2 \to Q_1}
  {\{P_1\} f \{Q_1\}}

\inferrule*[left=PreEx]
  {\forall x, \ \{P(x)\} f \{Q\}}
  {\{\lambda s . \ \exists x, P(x, s)\} f \{Q\}}

\inferrule*[left=Conj]
  {\{P\} f \{Q_1\} \\ \{P\} f \{Q_2\}}
  {\{P\} f \{Q_1 \land Q_2\}}
\end{mathpar}


We also prove rules for recursions and loops with break. The fixed-point rule for recursions formalizes the induction principle for the iterated function in their denotational semantics.
\begin{mathpar}
\inferrule*[left=Fix]
  {\forall W, \ (\forall a, \ \{P(a)\} W(a) \{Q\}) \to (\forall a, \ \{P(a)\} F(W, a) \{Q\})}
  {\forall a, \ \{P(a)\} \tf{Lfix}(F, a) \{Q\}}
\end{mathpar}

For loops with \lstinline{break}, we introduce two auxiliary operators to modularize proof.
One operator \lstinline{continue_case} assumes that \lstinline{x} has the form \lstinline{by_continue a} and unwraps it to return \lstinline{a}. The other one \lstinline{break_case} is analogous. These lead to a more modular repeat-break rule.
\begin{mathpar}
\inferrule*[left=RepeatBreak]
  {\forall a, \ \{P(a)\} x \gets f(a);; \tf{continue\_case}(x) \{P\} \\
   \forall a, \ \{P(a)\} x \gets f(a);; \tf{break\_case}(x) \{Q\}}
  {\forall a, \ \{P(a)\} \tf{repeat\_break}(f, a) \{Q\}}
\end{mathpar}


For set monad and state relation monad with errors, we build a similar Hoare logic as well. See Appendix \ref{appendixA} for more details.


\subsection{The Two-Stage Proof Approach}
Our framework introduces a two-stage proof approach to bridge the gap between natural reasoning and formal verification. This approach separates natural logical reasoning from mechanical composition, enhancing modularity and readability.

Consider for example, the \textit{match} procedure in the KMP algorithm shown in Algorithm \ref{match}. We formalize it using the set monad as follows. \lst{A} is the character type, and $\tf{range\_iter\_break}(l, h, f, j_0)$ is a for-loop with loop body being $f$, $i$ ranging from $l$ (included) to $h$ (excluded) and $j$ initialized as $j_0$.
\begin{lstlisting}
Context {A: Type} (default: A) 
        (patn text: list A) (next: list nat).
Definition inner_body(ch: A): nat -> program (ContinueOrBreak nat nat) :=
  fun j =>
    choice 
     (assume(ch = nth j patn default);; break (j+1))
     (assume(ch <> nth j patn default);;
      choice 
       (assume(j = 0);; break 0) 
       (assume(j <> 0);; continue (nth (j-1) next 0))).
Definition inner_loop(ch: A): nat -> program nat :=
  repeat_break (inner_body ch).
Definition match_body:
  nat -> nat -> program (ContinueOrBreak nat nat) :=
  fun i j =>
      let ch := nth i text default in
      j' <- inner_loop ch j;;
      choice
        (assume (j' = length patn);;
         break (i - length patn + 1))
        (assume (j' < length patn);;
         continue (j')).
Definition match_loop: program (ContinueOrBreak nat nat) :=
  range_iter_break 0 (length text) match_body 0.
\end{lstlisting}

The correctness of \textit{match} can be stated as follows: if $patn$ is nonempty, $patn$ and $next$ has the same length and $next$ is a prefix function of $patn$, then the return value $r$ is either $\textit{by\_break}(i)$ representing the first occurrence of $patn$ in $text$, or $\textit{by\_continue}(i)$ meaning there's no occurrence of $patn$ in $text$. We formalize it as the following Hoare triple:
\begin{gather*}
\left\{patn \neq nil \land patn.len = next.len \land \textit{prefix\_func}(next)\right\} \\
\tf{match\_loop} \\
\left\{\lambda r. \ \begin{cases}
    \textit{first\_occur}(i) & \text{if } r = \textit{by\_break}(i) \\
    \textit{no\_occur}(text.len) & \text{if } r = \textit{by\_continue}(i)
\end{cases} \right\}
\end{gather*}
The formal definition of these logical predicates and those introduced in Section \ref{sec::intro} can be found in Appendix \ref{appendixC}.

The first stage of our two-stage proof is the \textbf{essential proof} that captures the logical structure of the algorithm proof. In this stage, the proof proceeds by \textbf{logical groups}, each of which focuses on a logical topic and contains propositions relevant to the topic. For each group, we propose and prove \textbf{basic block propositions}, Hoare triples of relevant basic blocks. Tactics like \lst{hoare_auto} could facilitate the basic block verification. Between adjacent basic blocks, we prove \textbf{logical implications} connecting their preconditions and postconditions. These also include implications between pre/post-conditions of the procedure and pre/post-condition of some basic blocks.

We provide tactical support for decomposing programs into individual basic blocks and verifying them conveniently (see Appendix \ref{appendixB} for more details).
Luckily, for the \textit{match} example, the program is already well-structured so we do not need to transform it.
Then following the natural proof, our proof proceeds as follows:

\paragraph{Group 1: range.} In this group we prove \textit{jrange} holds throughout the for-loop as a foundation for other propositions. 
Since \textit{patn} is non-empty and \textit{next} has the same length as \textit{patn}, \textit{jrange} holds when entering the match loop: 
\[patn \neq nil \land patn.len = next.len \to jrange(0)\]
\textit{next} is a prefix function implies that its elements are within certain range.
\[\textit{prefix\_func}(next) \to \forall k: \textit{jrange}(k), next[k] \in [0, k] \]
By definition, basic block is a program segment without control flow transfer like loops and branches \cite{AhoLamSethiUllman:2007}. However, in our framework, it can contain \lstinline{choice} or even loops that have been verified previously, depending on the user's needs. Since \textit{jrange} is a simple proposition, we treat \lst{inner_body} as a basic block in this group.
When it continues, \textit{jrange} is preserved.
\begin{gather*}
\left\{ \textit{jrange}(j) \land (\forall k: \textit{jrange}(k), next[k] \in [0, k]) \right\} \\
x \gets \tf{inner\_body}(ch, j);; \tf{continue\_case}(x) \\ 
\left\{\lambda j'. \  \textit{jrange}(j') \right\}
\end{gather*}
When it breaks, \textit{jrange} no longer holds and the range of $j$ is as follows. 
\begin{gather*}
\left\{ \textit{jrange}(j)\right\}
x \gets \tf{inner\_body}(ch, j);; \tf{break\_case}(x) 
\left\{\lambda j'. \  j' \in [0, patn.len] \right\}
\end{gather*}
After exiting the inner loop, when outer loop continues, $jrange$ is back again.
\begin{gather*}
\left\{ j' \in [0, patn.len] \right\}
\tf{assume}(j' < patn.len);; \tf{ret}(j')
\left\{\lambda j''. \  \textit{jrange}(j'') \right\}
\end{gather*}
Some reader may point out that in the original program it ends with $\tf{continue}(j')$ instead of  $\tf{ret}(j')$. This does not matter since $x \gets \tf{continue}(j');; \tf{continue\_case}(x)$ is equivalent to $\tf{ret}(j')$. When applying the repeat-break rule in the second stage, the original basic block will become just like this.

\paragraph{Group 2: partial match.} In this group we prove that $\textit{partial\_match}(i, j)$ is an invariant of the for-loop, but we do not care about whether $j$ is the best result. Obviously, it holds when the outer loop begins:
$\textit{partial\_match}(0, 0)$.
Based on \textit{jrange} and preconditions, $\textit{partial\_match}(i, j)$ is preserved by the continue branch of the inner body.
\begin{gather*}
\left\{ next.len \leq patn.len \land \textit{prefix\_func}(next) \land \textit{jrange}(j) \land  \textit{partial\_match}(i, j) \right\} \\
\tf{assume}(j \neq 0);; \tf{ret}(next[j-1]) \\ 
\left\{\lambda j'. \  \textit{partial\_match}(i, j') \right\}
\end{gather*}
When inner loop breaks, we aim to prove the original $\textit{partial\_match}(i, j)$ is extended to next $i$ and current $j$, i.e. $\textit{partial\_match}(i+1, j')$.
The first case where $text[i] = patn[j]$ needs preconditions and the range of $i$ in the for-loop.
\begin{gather*}
\left\{ next.len \leq patn.len \land i \in [0, text.len) \land \textit{jrange}(j) \land \textit{partial\_match}(i, j) \right\} \\
\tf{assume}(text[i] = patn[j]);; \tf{ret}(j+1) \\ 
\left\{\lambda j'. \  \textit{partial\_match}(i+1, j') \right\}
\end{gather*}
The second case where $j=0$ is trivial.
\begin{gather*}
\left\{ \right\}
\tf{assume}(j=0);; \tf{ret}(0) 
\left\{\lambda j'. \  \textit{partial\_match}(i+1, j') \right\}
\end{gather*}
$\textit{partial\_match}(i+1, j')$ is preserved when the for-loop continues because $j'$ is not changed.

\paragraph{Group 3: partial bound.} In this group, to prepare for the two \textit{no\_occur} in the postcondition, we prove that \textit{partial\_bound} is an invariant of the for-loop.
It holds when the for-loop starts:
$\textit{partial\_bound}(0, 0)$.
For inner loop, we propose a new invariant $\textit{presuffix\_inv}(i, j)$.
The two invariants of the for-loop, $\textit{partial\_match}(i, j)$ and $\textit{partial\_bound}(i, j)$, ensure $\textit{presuffix\_inv}(i, j)$ when entering the inner loop.
\begin{gather*}
next.len \leq patn.len \land  i \in [0, text.len) \land \textit{partial\_match}(i, j) \land \\ \textit{partial\_bound}(i, j)  \to \textit{presuffix\_inv}(i, j) 
\end{gather*}
When the inner loop continues, $\textit{presuffix\_inv}(i, j)$ is preserved.
\begin{gather*}
\{next.len \leq patn.len \land \textit{prefix\_func}(next) \land \textit{jrange}(j) \land \\ 
\textit{presuffix\_inv}(i, j) \land text[i] \neq patn[j] \} \\
\tf{assume}(j \neq 0);; \tf{ret}(next[j-1]) \\ 
\left\{\lambda j'. \  \textit{presuffix\_inv}(i, j') \right\}
\end{gather*}
When inner loop breaks, we leverage $\textit{presuffix\_inv}(i, j)$ to prove $\textit{partial\_bound}(i+1, j')$.
The first case is $text[i] = patn[j]$:
\begin{gather*}
\left\{ next.len \leq patn.len \land \textit{jrange}(j) \land \textit{presuffix\_inv}(i, j) \right\} \\
\tf{assume}(text[i] = patn[j]);; \tf{ret}(j+1) \\ 
\left\{\lambda j'. \  \textit{partial\_bound}(i+1, j') \right\}
\end{gather*}
The second case is $text[i] \neq patn[j]$:
\begin{gather*}
\left\{ text[i] \neq patn[j] \land \textit{presuffix\_inv}(i, j) \right\} \\
\tf{assume}(j=0);; \tf{ret}(0) \\ 
\left\{\lambda j'. \  \textit{partial\_bound}(i+1, j') \right\}
\end{gather*}
Similar to group 2, it is preserved when for-loop continues.

\paragraph{Group 4: post loop.} In this group we prove the postcondition using the previous results. Firstly, we prove that $\textit{no\_occur}(i)$ is an invariant of for-loop. Obviously $\textit{no\_occur}(0)$. Based on $\textit{partial\_bound}(i+1, j')$ and $\textit{jrange}(j')$, we can prove it is preserved.
\begin{gather*}
i \in [0, text.len) \land \textit{jrange}(j') \land \textit{partial\_bound}(i+1, j') \land \textit{no\_occur}(i) \to \\
\textit{no\_occur}(i+1)
\end{gather*}
If the for-loop terminates because $i$ reaches the upperbound, then we have $\textit{no\_occur}(text.len)$, which is half of our postcondition.
On the other hand, if the for-loop terminates because it breaks, we can prove the \textit{no\_occur} part of \textit{first\_occur}. 
\begin{gather*}
\left\{ \textit{no\_occur}(i) \right\} \\
\tf{assume}(j' = patn.len);; \tf{ret}(i - patn.len + 1) \\ 
\left\{\lambda r. \  \textit{no\_occur}(r + patn.len - 1) \right\}
\end{gather*}
Using $\textit{partial\_match}(i+1, j')$, we can prove the other part of \textit{first\_occur}.
\begin{gather*}
\left\{ \textit{partial\_match}(i+1, j') \right\} \\
\tf{assume}(j' = patn.len);; \tf{ret}(i - patn.len + 1) \\ 
\left\{\lambda r. \  text[r..r+patn.len] = patn \right\}
\end{gather*}

Although the proof is complete from natural understanding, the formal proof needs to establish an end-to-end correctness theorem. Therefore, the second stage is the \textbf{mechanized proof}, which composes the essential proof’s results into a complete formal argument using Hoare rules.
In this stage, by repeatedly using conjunction rule and consequence rule, we can merge grouped propositions of the same basic block into a single Hoare triple. Then we can use choice rules to combine two branches' Hoare triples, use loop rules to transform the loop body's Hoare triple into the loop's Hoare triple, and use bind rule to combine them altogether.

In our example, combining results from the essential proof, the inner body satisfy following Hoare triples:
\begin{gather*}
\{next.len \leq patn.len \land \textit{prefix\_func}(next) \land (\forall k : \textit{jrange}(k), next[k] \in [0, k]) \land \\
\textit{jrange}(j) \land \textit{partial\_match}(i, j) \land \textit{presuffix\_inv}(i, j) \} \\
x \gets \tf{inner\_body}(text[i], j);; \tf{continue\_case}(x) \\ 
\left\{\lambda j'. \  \textit{jrange}(j') \land \textit{partial\_match}(i+1, j') \land \textit{presuffix\_inv(i+1, j')} \right\}
\end{gather*}
\begin{gather*}
\{next.len \leq patn.len \land \textit{prefix\_func}(next) \land i \in [0, text.len) \land \\
\textit{jrange}(j) \land \textit{partial\_match}(i, j) \land \textit{presuffix\_inv}(i, j) \} \\
x \gets \tf{inner\_body}(text[i], j);; \tf{break\_case}(x) \\ 
\left\{\lambda j'. \ j' \in [0, patn.len] \land \textit{partial\_match}(i+1, j') \land \textit{partial\_bound(i+1, j')} \right\}
\end{gather*}
Using the repeat-break rule and the consequence rule, we further obtain the Hoare triple of the inner loop:
\begin{gather*}
\{next.len \leq patn.len \land \textit{prefix\_func}(next) \land i \in [0, text.len) \land \\
\textit{jrange}(j) \land \textit{partial\_match}(i, j) \land \textit{partial\_bound}(i, j) \land \textit{no\_occur}(i) \} \\
\tf{inner\_loop}(text[i], j) \\ 
\{\lambda j'. \  j' \in [0, patn.len] \land \textit{partial\_match}(i+1, j') \land \\
\textit{partial\_bound}(i+1, j') \land \textit{no\_occur}(i+1) \}
\end{gather*}
Then similarly we can obtain Hoare triples for the match body and the match loop, finishing the proof.

This two-stage proof approach improves modularity by splitting complicated propositions of complex programs into simple propositions of basic blocks. Besides, it enhances readability by ensuring that proofs reflect the natural logical structure of the correctness argument rather than the program’s syntactic structure. Moreover, the approach achieves generality, as it can be applied to almost all formal frameworks based on Hoare logic. 

In addition, we also formalize and prove the table-building procedure in the KMP algorithm, further showcasing the usability of our framework.

%% file: related.tex
\section{Related Work}\label{sec::related}
 The Isabelle Refinement Framework \cite{Refine_Monadic-AFP} provides a monadic approach to program verification, enabling users to specify and refine nondeterministic algorithms in a functional style. While it supports abstract specifications and stepwise refinement, it is primarily designed for functional and stateless programs, making it less suitable for imperative algorithms with complex state transitions. In contrast, our framework introduces customizable states and more flexible control flow constructs such as loops with break, which are convenient for naturally specifying algorithms like DFS and KMP. Additionally, it focuses on the data refinement and program refinement, but our work focuses more on specifying and verifying abstract algorithms.

 Separation Logic \cite{reynolds2002separation} provides a foundation for reasoning about programs with mutable state and pointers, focusing on local reasoning about memory. Frameworks like the Iris \cite{iris} build upon Separation Logic to verify concrete programs, often involving complex memory manipulations. While these frameworks excel at verifying low-level implementation details, our approach focuses on specifying and verifying algorithms at a higher level of abstraction using a state relation monad, aiming to separate the core algorithmic logic from more concrete implementation concerns like memory management.

Nigron et al. also developed a framework with a Hoare logic for monadic programs in Coq \cite{Nigron:2021}. While their work, like ours, utilizes monads and builds a corresponding Hoare logic, the motivations and resulting logics differ significantly. They aimed to reason about identifier freshness generated by monadic programs. To achieve this, they employed separation logic principles, tailoring their Hoare logic to enable local reasoning about freshness, notably featuring the frame rule. In contrast, our framework uses the state relation monad primarily to specify algorithms naturally and concisely. Consequently, our Hoare logic is designed to mirror the structure of natural proofs, featuring rules for various monadic operators, loop constructs, and recursion, alongside structural rules like the conjunction rule to facilitate modular natural reasoning about algorithm correctness.

Lammich and Neumann's framework for verifying DFS algorithms \cite{DFS-Framework} provides a structured approach to specifying and proving the correctness of DFS using a combination of refinement techniques and modular proof components. Since it is built on the Isabelle Refinement Framework, it represents program states as explicit arguments. In contrast, our framework abstracts away the state using a state relation monad, enabling more concise and elegant formulations of algorithms. Besides, while the DFS framework features the design principle and technique of incrementally establishing invariants, our two-stage proof approach can not only achieve the same effect, but also handle more complex structures, such as multiple layers of loops and intricate branches.
This makes our framework suitable for a wider range of algorithms.

There has been formal proof of the KMP algorithm, such as Paulson's \cite{KMP-AFP}. It describes the table-building and matching procedures as single-layer loops and uses traditional Hoare logic for verification. In contrast, we model both procedures as two-layer loops that share the same inner loop, aligning more closely with common practices. Our proof framework allows us to provide a modular and readable proof for such formulation.

%% file: conclucion.tex
\section{Conclusion}\label{sec::conclu}
In this paper, we introduced a formal framework for naturally specifying and verifying sequential algorithms in Coq. Our approach leverages a state relation monad to integrate Coq's expressive type system with the flexible control flow of imperative languages. This allows for the specification of algorithms at an appropriate level of abstraction, supporting nondeterministic operations and customizable program states. We provided stateless and errorful variants of the monad. We also developed a proof framework for partial correctness tailored to our monad, which includes Hoare rules for various statements and a novel two-stage proof approach. This approach separates natural logical reasoning from mechanical composition, enhancing modularity and readability. We demonstrated the versatility and applicability of our framework through practical evaluations, including the formalization of the DFS algorithm and the verification of the KMP algorithm.


%% file: appendix.tex
\section{Stateless and Errorful Variants of the State Relation Monad}\label{appendixA}
\subsection{Set Monad}
For some algorithms, it is unnecessary to involve states. Although we can use a state relation monad with \lstinline{unit} state (as we've done in previous examples), it is more convenient and natural to use a monad that is intrinsically stateless. Therefore, we provide a stateless variant of our monad: the \textit{set monad}.

A set monad is a set, representing possible results of a monadic program.
Its type constructor is \lstinline{program (A: Type) := A -> Prop}, with \lstinline{ret} and \lstinline{bind} defined as follows:
\[\begin{aligned} 
  r \in \texttt{ret}(a) &\iff r = a \\
   b \in \texttt{bind}(c, f) &\iff \exists\ a, \ a \in c \ \land \ b \in f(a)
\end{aligned}\]
Other operators are defined as stateless versions of those in state relation monad. Since \lstinline{assume} is already stateless we do not need \lstinline{assume'} for the set monad.

For set monad, since the precondition becomes stateless and can be introduced as a logical premise, a Hoare triple degenerates into a Hoare pair: 
\begin{lstlisting}
Definition Hoare (c: program A) (Q: A -> Prop): Prop :=
  forall a, a inx c -> Q a.
\end{lstlisting}
Nevertheless, a set monad program usually has some arguments and premises are about the arguments, then its proposition has the form $P(a) \to \tf{Hoare}(c(a), Q)$, which can be regarded as a Hoare triple $\{P(a)\}c(a)\{Q\}$. Hence, proof rules for Hoare triples can be easily adopted for Hoare pairs and we can also use the notation of Hoare triples ($\{P\}c\{Q\}$) for Hoare logic on set monad.

\subsection{State Relation Monad with Errors}
In practical algorithms, it is often necessary to model errorful computations. To support this, we propose state relation monad with errors. It extends the state relation monad to a Coq \lstinline{record} with two fields: \lstinline{nrm} and \lstinline{err}. The \lstinline{nrm} field represents the normal execution of a program, defined identically to the state relation monad. The \lstinline{err} field represents errorful executions, defined as the set of states that result in an error.
\begin{lstlisting}
Record program (sigma A: Type): Type := {
  nrm: sigma -> A -> sigma -> Prop;
  err: sigma -> Prop
}.
\end{lstlisting}

\lstinline{ret} returns a value with no error. 
\[\begin{aligned}
   (s_1, r, s_2) \in (\tf{ret}(a)).\tf{nrm} &\iff r = a \ \land \ s_1 = s_2 \\
   (\tf{ret}(a)).\tf{err} & \ = \ \emptyset
\end{aligned}
\]

The \lstinline{bind} operation composes two programs, with $(\tf{bind}(c,f)).\tf{nrm}$ representing the case where both $c$ and $f(a)$ execute normally, and $(\tf{bind}(c,f)).\tf{err}$ capturing cases where either $c$ results in an error or $c$ executes normally but $f(a)$ results in an error:
\[\begin{aligned}
  & (s_1, b, s_3) \in (\tf{bind}(c,f)).\tf{nrm} \\ \iff 
  &\exists \ a \ s_2, \ (s_1, a, s_2) \in c.\tf{nrm} \ \land \ (s_2, b, s_3) \in f(a).\tf{nrm} \\
  & s \in \tf{bind}(c,f)).\tf{err} \\ \iff 
  & s \in c.\tf{err} \ \lor \ (\exists \ a \ s_2, \ (s_1, a, s_2) \in c.\tf{nrm} \ \land \ s_2 \in f(a).\tf{err})
\end{aligned}
\]

The \lstinline{choice} operation takes the union of both the \lstinline{nrm} and \lstinline{err} fields of two programs. Other operators, such as \lstinline{assume} and \lstinline{update}, are extended with \lstinline{err}set to $\emptyset$. Additionally, we introduce a new operator, \lstinline{assert}, which asserts a proposition about the program state and produces an error if the proposition does not hold.

For state relation monad with errors, the Hoare triple states that a program is not only functionally correct when it executes normally but also error-free.
\begin{lstlisting}
Definition Hoare (P:sigma -> Prop) (c: program sigma A) (Q: A -> sigma -> Prop) :=
  (forall s1 a s2, P s1 -> (s1, a, s2) inx c.(nrm) -> Q a s2)
  /\ (forall s1, P s1 -> s1 notinx c.(err)).
\end{lstlisting}
Most rules for state relational monad can be transferred without major change.

\section{Support for Basic Block Verification}\label{appendixB}
Our proof approach involves decomposing programs into basic blocks and verifying them independently. We provide tactics to automate them and reduce user effort.

When splitting a program, we may have to transform its syntactic structure.
For example, consider the following program:
\begin{lstlisting}
Example c1 (f g: program sigma Z) (p q: sigma -> Prop) (h: Z -> Z -> program sigma Z)
  := x <- f;; assume(p);; y <- g;; assume(q);; h x y.
\end{lstlisting}
If we want to isolate the basic block before \lst{h x y} as a standalone program, directly picking the first four statements does not work. This is because the basic block ends with \lst{assume} and returns a unit type, thus the value of \lst{x} and \lst{y} cannot be passed to the subsequent program \lst{h}.
To address this problem, we can transform \lst{c1} into an equivalent form:
\begin{lstlisting}
Example c2 (f g: program sigma Z) (p q: sigma -> Prop) (h: Z -> Z -> program sigma Z)
  := '(x, y) <- (x <- f;; assume(p);; y <- g;; assume(q);; ret (x, y));; 
     h x y.
\end{lstlisting}
Then it is easy to split the block and \lst{h} apart.
To prove \lst{c1} and \lst{c2} are equivalent, we repeatedly apply monad laws and a lemma \lst{bind_equiv} that allows us to prove $\tf{bind}(c_0, f_0) = \tf{bind}(c_1, f_1)$ by proving $c_0 = c_1$ and $f_0 = f_1$.
We provide a tactic \lst{monad_laws} to repeatedly apply monad laws and a tactic \lst{monad_equiv} to prove syntactical program equivalence using this method.

Apart from splitting programs apart, we also provide a tactic \lst{hoare_auto} to facilitate verification of each basic block. This tactic leverages lemmas including monad laws to rewrite the program into a proper form, applies suitable Hoare rules based on the syntactic structure of the Hoare triple and finally obains verification conditions (VCs) as proof goals. This tactic can also deal with programs with \lst{choice} besides basic blocks.

For instance, consider the following Hoare triple, where \lst{hailstone_body} is the loop body of the program \lst{hailstone} mentioned before.
\[
\{ \lambda s. \ x \geq 1\} 
 y \gets \tf{hailstone\_body}(x);; \tf{continue\_case}(y)
\{ \lambda ys. \ y \geq 1\}
\]
\lst{hoare_auto} will reduce the triple to the following proof goals:
\begin{gather*}
\text{goal 1: }(\exists k, x = 2k) \land x \geq 1 \implies \frac{x}{2} \geq 1 \\
\text{goal 2: }(\exists k, k \neq 0 \land x = 2k+1) \land x \geq 1 \implies 3x+1 \geq 1
\end{gather*}

\section{Formal Definition of Logical Predicates}\label{appendixC}
The formal definition of key logical predicates in the KMP match procedure is as follows:
\begin{itemize}
    \item $\textit{jrange}(j)$: \\$0 \leq j < next.len$.
    \item $\textit{partial\_match}(i, j)$: \\$patn[0..j] = text[i-j..i]$.
    \item $\textit{presuffix}(a, b)$: \\$patn[0..a]$ is both a prefix and a suffix of $patn[0..b]$.
    \item $\textit{proper\_presuffix}(a, b)$: \\$\textit{presuffix}(a, b)$ and $a < b$.
    \item $\textit{presuffix\_bound}(a,b)$: \\for any $c$, if $\textit{proper\_presuffix}(c, b)$, then $\textit{presuffix}(c, a)$.
    \item $\textit{prefix\_func}(next)$: \\for any $i \in [0, next.len)$, $\textit{proper\_presuffix}(next[i], i+1) \land \textit{presuffix\_bound}(next[i], i+1)$.
    \item $\textit{no\_occur}(i)$: \\for any $j \in [0, i-patn.len]$, $text[j..j+patn.len] \neq patn$.
    \item $\textit{first\_occur}(i)$: \\$text[i..i+patn.len] = patn$ and $\textit{no\_occur}(i+patn.len-1)$. 
    \item $\textit{partial\_bound}(i, j)$: \\for any $z$, if $\textit{partial\_match}(i, z)$ then $z \leq j$.
    \item $\textit{presuffix\_inv}(i, j)$: \\for any $k > 0$, if $\textit{partial\_match}(i+1, k)$ then $\textit{presuffix}(k-1, j)$ and $patn[k-1] = text[i]$.
\end{itemize}